\documentclass[a4paper,11pt]{article}
\pdfoutput=1 

\usepackage{jheppub} 

\usepackage[T1]{fontenc} 

\title{\boldmath Black holes in 4D Einstein-Maxwell-Gauss-Bonnet gravity coupled with scalar fields}
\author[a]{Yi-Li Wang,}
\author[b,c]{Xian-Hui Ge}

\affiliation[a]{Arnold Sommerfeld Center for Theoretical Physics, Ludwig-Maximilians-Universit\"at M\"unchen, Germany}
\affiliation[b]{Department of Physics, Shanghai University, P.R.China}
\affiliation[c]{Center for Gravitation and Cosmology, College of Physical Science and Technology, Yangzhou University, P.R.China}

\emailAdd{gexh@shu.edu.cn}
\emailAdd{Wang.Yili@physik.uni-muenchen.de}

\abstract{Einstein-Maxwell-Gauss-Bonnet-axion theory in $4$-dimensional spacetime is investigated in this paper through a ``Kaluza-Klein-like'' process. Dual to systems at finite temperature with background magnetic field on three dimensions, the four-dimensional dyonic black hole solution coupled with higher derivative terms is obtained. After the tensor-type perturbation is added, the shear viscosity to entropy density ratio is calculated at high temperature and low temperature separately. The behaviour of shear viscosity to entropy density ratio of uncharged black holes is found to be similar with that in $5$-dimensional spacetime,  violating the Kovtun-Starinets-Son bound as well when temperature becomes lower. In addition, the main feature of this ratio remains almost unchanged in $4$ dimensions, which is characterised by $(T/\Delta)^2$ at low temperature $T$, with $\Delta$ proportional to the coefficient $\beta$ from scalar fields. The difficulty in causal analysis is also discussed, which is mainly caused by the vanishing momentum term in equations of motion.}

\begin{document} 
\maketitle
\flushbottom

\section{Introduction}
Lovelock's theory suggests that Einstein gravity can be modified with higher derivative terms, with second order equations of motion \cite{ll1,ll2}. One example is the well-known Einstein-Gauss-Bonnet (EGB) gravity. Increasing interest has been put on this sort of gravity in $4$-dimensional spacetime. Recent research \cite{lcs} presents a method to realise it through rescaling Gauss-Bonnet coupling constant $\tilde{\alpha}\to\alpha/(D-4)$, and taking $D\to4$ to obtain spherically symmetric 4D black hole solutions with non-vanishing Gauss-Bonnet term.\\
The strategy is quite straightforward. Considering an action with the contribution from the Gauss-Bonnet term after the rescaling of Gauss-Bonnet constant $\tilde{\alpha}$, one has
\begin{eqnarray}
S&=&\frac{1}{16\pi G_D}\int d^Dx\sqrt{-g}\bigg(R-2\Lambda\left.+\frac{\alpha}{D-4}(R_{\mu\nu\rho\sigma}R^{\mu\nu\rho\sigma}-4R_{\mu\nu}R^{\mu\nu}+R^2)\right).
\end{eqnarray}
The action then yields the equations of motion:
\begin{equation}
R_{\mu\nu}-\frac 12 g_{\mu\nu}+\Lambda g_{\mu\nu}+\frac{\alpha}{D-4}H_{\mu\nu}=0,
\end{equation}
where $H_{\mu\nu}$ is the Gauss-Bonnet tensor
\begin{eqnarray}
H_{\mu\nu}&=&2\left[RR_{\mu\nu}-2R_{\mu\rho\nu\sigma}R^{\rho\sigma}+R_{\mu\alpha\beta\gamma}R_{\nu}^{\ \alpha\beta\gamma}-2R_{\mu\alpha}R^{\alpha}_{\ \nu}\right.\nonumber\\
&&\left.-\frac 14g_{\mu\nu}(R_{\mu\nu\rho\sigma}R^{\mu\nu\rho\sigma}-4R_{\mu\nu}R^{\mu\nu}+R^2)\right].
\end{eqnarray}
As we know, $H_{\mu\nu}$ vanishes in $4$ dimensions and the theory will reduce to Einstein's gravity. After being rescaled, however, it seems that the infinity caused by $(D-4)$ will leave us a non-vanishing term. So it was suggested that taking $D\to4$ limit will give Gauss-Bonnet gravity in $4$ dimensions:
\begin{equation}
\lim\limits_{D\to 4}\left[R_{\mu\nu}-\frac 12 g_{\mu\nu}+\Lambda g_{\mu\nu}+\frac{\alpha}{D-4}H_{\mu\nu}\right]=0.
\end{equation}
This work sheds a light on the investigation into higher derivative gravity in four-dimensional spacetime. However, much research implies that the strategy performed \cite{lcs} actually cannot provides Gauss-Bonnet gravity in four dimensions \cite{ar1,ar2,ar3,ar4,ar5,ar6,ar7}, i.e. it is not a ``novel Einstein-Gauss-Bonnet gravity''.

Since the strategy proposed above is unable to give topologically non-trivial solutions, another method is supposed with ``Kaluza-Klein-like'' procedure \cite{2003,galieon}, compactifying D-dimensional EGB gravity which is on a maximally symmetric $(D-4)$-dimensional space. With similar rescaling and $D\to4$ limit, one obtains a purely 4-dimensional EGB theory \cite{2003,2004,2006}. The resulting theory can also be viewed as a special Hondeski gravity or generalised Galileons \cite{2003,hondeski,gali1,gali2}.

Meanwhile, as black holes have a good feature of thermodynamical properties \cite{hawking}, great attention has been paid on this field.  During the past decades, the AdS/CFT dictionary provides a powerful tool to investigate strongly coupled gauge theories at finite temperature in $D$ dimensions, which are dual to black holes (or black branes) in $(D+1)-$ dimensional AdS space. The temperature of these systems is equal to the Hawking temperature of the dual black hole. The thermodynamical properties of the black hole can characterise those of the dual gauge theories, and the thermal phase structure of the system as well. 

With the recently developed $4D$ Gauss-Bonnet gravity, one is capable of investigating deeper on $(2+1)-$ dimensional strongly interacting systems with implications from the gauge/gravity duality. The existence of higher derivative terms corresponds to 't Hooft coupling constant corrections in the boundary theories. One main goal in this paper is to build a new system with such corrections through the holographic method, i.e. deriving a new black hole solution corresponding to it. This new solution aims to bring a physical insight into the high-temperature superconductivity in the future.

One of the fundamental methods to study conducting medium is to put it under a magnetic field and find its behaviour. Therefore, the system to be constructed here would be more useful for the future research if a background magnetic field is considered. Hence, to build a $(2+1)D$ model at a finite charge density which duals to Gauss-Bonnet gravity with a background magnetic field, we will add both electric and magnetic charges to the black hole.

One also notes that the realistic materials do not necessarily preserve the spatial translation invariance. To explore a more realistic behaviour of the systems, one needs to break this symmetry. A simple way to obtain the momentum relaxation is to consider a theory coupling with spatial scalar fields \cite{axion}. The behaviour of systems with scalar fields have been studied widely \cite{axion,scalar1,scalar2,scalar3,scalar4,scalar5,scalar6,scalar7,scalar8,scalar9,scalar10,scalar11,scalar12,scalar13,scalar14}. An additional advantage of coupling with scalar fields can be seen from the DC conductivity. The DC conductivity obtained by perturbing the boundary with a field of frequency $\omega$ and taking $\omega\to 0$ \cite{dc1,dc2} will blow up to infinity if the momentum is conserved. The scalar fields bring momentum dissipation and enable us to work with a theory with finite DC conductivity. So in order to construct a more realistic model, building a solid foundation for our future investigation in its properties, we will add scalar fields to break the translational symmetry as well. 

After deriving the new black hole solution, we will focus on the shear viscosity to entropy density ratio $\eta/s$ of it, which is a feature of great importance. This ratio tells people to which extend is a given fluid to be ``perfect''. Exploring $\eta/s$ in higher derivative gravity, can also help to rule out some of the corrections in gravitational theories. For gauge theories, there exists a universal bound, known as the Kovtun-Starinets-Son (KSS) bound for shear viscosity to entropy density ratio \cite{kss1,kss2,kss3,kss4,kss5}:

\begin{equation}\label{kss}
\frac{\eta}{s}\geq\frac{1}{4\pi}.
\end{equation}

This bound is saturated for the gauge theories dual to classical Einstein gravity. It provides a powerful tool for estimating the viscosity for strongly coupled systems. But the KSS bound would be violated when small corrections are added to Einstein's gravity. Coupled with Gauss-Bonnet term, a modified version of this bound in $5D$ spacetime reads \cite{GBV}
\begin{equation}
\frac{\eta}{s}\geq\frac{1}{4\pi}(1-4{\alpha}_{GB}),
\end{equation} 
where $\alpha_{GB}=-\Lambda\alpha/3$, with $\Lambda$ the cosmological constant. Extended to $D-$dimensional spacetime, this bound reads \cite{4d1}

\begin{equation}
\frac{\eta}{s}\geq\frac{1}{4\pi}\left(1-\frac{2(D-1)}{D-3}\alpha_{GB}\right).
\end{equation}
Consequently, in $4D$ cases,
\begin{equation}\label{gbbound}
\frac{\eta}{s}\geq\frac{1}{4\pi}\left(1-6\alpha_{GB}\right).
\end{equation}

It has been shown that when non-vanishing electric charge is involved, bound (\ref{gbbound}) will be violated in $4$D Gauss-Bonnet gravity, and constraint on Gauss-Bonnet coupling could be obtained by analysing the causal structure in the bulk \cite{4d1}.

In this paper, it is of our interest to check whether violation will happen and will there be a new constraint for the coupling constant if we include two scalar fields linear in all the spatial directions.

In this paper, $4$D black hole solution coupled with higher derivative terms with both electric and magnetic charges as well as axions will be obtained. The ``Kaluza-Klein-like'' procedure introduced in \cite{2003} will be used to find the solution. Then, the shear viscosity to entropy density ratio for a neutral black hole with scalar fields will be studied, which will be found to violate the KSS bound.

This paper is organised as follows:

In section $2$, the Einstein-Maxwell-Gauss-Bonnet-axion gravity will be studied through ``Kaluza-Klein-like'' method, and a dyonic black hole solution coupled with scalar fields and higher derivative terms will be obtained.

Removing electric and magnetic charges, in section $3$, we will calculate the shear viscosity to entropy density ratio for the neutral black holes in $4$D Gauss-Bonnet gravity with axions. It will be shown that the KSS bound will be violated because of the scalar fields.

Section $4$ will focus on the difficulties one may meet when finding constraints on Gauss-Bonnet constant through causality analysis.

Finally, conclusion and outlook will be represented in section $5$.

\section{Dyonic black holes in four dimensions with Gauss-Bonnet coupling}
\subsection{Reduced action}
Working in four dimensions, we first introduce two scalar fields. The general action in $D$-dimensional spacetime for Einstein-Maxwell-Gauss-Bonnet-axion (EMGBA) gravity reads
\begin{eqnarray}\label{ori_action}
S&=&\frac{1}{16\pi G_D}\int d^Dx\sqrt{-g}\left(R-2\Lambda+\tilde{\alpha}\mathcal{L}_{GB}-\frac 12\sum_{i=1}^{i=2}(\partial \varphi_i)^2-\frac 14 F_{\mu\nu}F^{\mu\nu}\right),
\end{eqnarray}
where $\Lambda=-(D-1)(D-2)/2l^2$ is the cosmological constant, and $l$ represents the AdS radius. The Gauss-Bonnet term $\mathcal{L}_{GB}$ takes the form
\begin{equation}
\mathcal{L}_{GB}=R_{\mu\nu\rho\sigma}R^{\mu\nu\rho\sigma}-4R_{\mu\nu}R^{\mu\nu}+R^2.
\end{equation}

The scalar fields $\varphi_i$ in this action are taken to be massless, in order that their contribution in bulk energy-stress tensor will only be $\partial_{\mu}\varphi_i$. On the other hand, two such fields are added, because the $4D$ gravity will be studied, and it corresponds to the $3D$ boundary theory, whose spatial dimension is $2$. With the same number of the scalar fields, making these fields linear in spatial coordinates, one can obtain an isotropic bulk solution.

The field strength enters the action as $F_{\mu\nu}=\partial_{\mu}A_{\nu}-\partial_{\nu}A_{\mu}$ for a $U(1)$ gauge field $A$ carrying a electric charge $q$ and a magnetic charge $h$.

Now we parameterise a $D$-dimensional metric:
\begin{equation}
ds_D^2=ds_p^2+e^{2\phi}d\Sigma_{D-p,\lambda}^2,
\end{equation}
where $\phi$ is called the ``breathing scalar'', depending only on external p-dimensional coordinates \cite{galieon,2003,2004}. The line elements $d\Sigma_{D-p,\lambda}^2$ describe the internal maximally symmetric space, and $\lambda$ relates to the curvature of the internal spacetime. This is a diagonal reduction along $\Sigma_{D-p,\lambda}$. When $\lambda=0$, the ``internal'' space is flat. Here we only consider Abelian isometry group for the ``internal'' space, so the massive modes could be truncated.\\
Now let us rescale Gauss-Bonnet coupling constant, setting 
\begin{equation}
\alpha=\epsilon \tilde{\alpha},
\end{equation}
where $\epsilon=(D-p)$. The next step is to consider the cases where $p<5$, and take the limit $D\to p$. The resulting reduced $p$-dimensional action reads \cite{galieon,2003}
\begin{eqnarray}\label{action}
S_{{reduced}}&=&\frac{1}{16\pi G_p}\int d^px\sqrt{-g}\big[R-2\Lambda
+\alpha\left(\phi\mathcal{L}_{GB}+4G^{\mu\nu}\partial_{\mu}\phi\partial_{\nu}\phi-4(\partial\phi)^2\square\phi+2((\partial\phi)^2)^2\right)\nonumber\\
&&-\frac 12\sum_{i=1}^2(\partial\varphi_i)^2-\frac 14F_{\mu\nu}F^{\mu\nu}\left.-2\lambda R e^{-2\phi}-12\lambda(\partial\phi)^2e^{-2\phi}-6\lambda^2e^{-4\phi}\right],
\end{eqnarray}
where $G_{\mu\nu}$ are Einstein tensors. This action is the EMGBA theory in $4$ dimensions, and it also works for $D<4$ \cite{2003,2004}. Constructed in a mathematically more rigorous way, (\ref{action}) does not suffer from defects of the naive $D\to4$ limit such as being ill-defined \cite{ar1,ar2,2004}. 

\subsection{Dyonic Black hole solution}
The next step is to find the solution with planar symmetry. One could first get equations of motion from the action (\ref{action}) in four dimensions ($p=4$). There are four sets of equations. The first are the Klein-Gordon equations:
\begin{equation}
\mathcal{E}_{\varphi}=\nabla_{\mu}\nabla^{\mu}\varphi_i=0.
\end{equation}

These equations are naturally satisfied by choosing spatially linear fields $\varphi_i=\beta x_i$, where $\beta$ is a constant, and $x_i=\{x,y\}$. Then are the Maxwell equations which read
\begin{equation}
\mathcal{E}_F=\nabla_{\mu}F^{\mu\nu}=0.
\end{equation}
The gauge field is assumed to be \cite{charge1}
\begin{equation}
A=A_t(r)dt+A_y(x)dy.
\end{equation}
The solution of these equations is
\begin{equation}
A=(A_t(r),0,0,A_y(x))=(-\frac qr,0,0,hx),
\end{equation}
where the electric charge $q$ enters in the $t-$ component, while the magnetic charge $h$ is in the $y-$ direction. So the black hole carries these charges with
\begin{equation}
F=\frac{q}{r^2}dr\wedge dt+hdx\wedge dy.
\end{equation}

The variation with respect to the ``breathing scalar'' $\phi$ yields the equation \cite{2004}
\begin{eqnarray}
\mathcal{E}_{\phi}&=&-\mathcal{L}_{GB}+8G^{\mu\nu}\nabla_{\nu}\nabla_{\mu}\phi+8R^{\mu\nu}\nabla_{\mu}\phi\nabla_{\nu}\nonumber\\
&&-8(\square\phi)^2+8(\nabla\phi)^2\square\phi+16\nabla^{\mu}\phi\nabla^{\nu}\nabla_{\nu}\nabla_{\mu}\phi\nonumber\\
&&+8\nabla_{\mu}\nabla_{\nu}\nabla^{\mu}\nabla^{\nu}\phi-24\lambda^2e^{-4\phi}-4\lambda Re^{-2\phi}\nonumber\\
&&+24\lambda e^{-2\phi}\left((\nabla\phi)^2-\square\phi\right)=0,
\end{eqnarray}
where $G_{\mu\nu}$ are Einstein tensors. Finally is the Einstein equation \cite{2004}
\begin{eqnarray}
{\mathcal E}_{\mu\nu} &=& \Lambda g_{\mu\nu} +  G_{\mu\nu}-\sum_{i=1}^{2}\left(\frac 12\partial_{\mu}\varphi_i\partial_{\nu}\varphi_i-\frac{g_{\mu\nu}}{4}(\partial \varphi_i)^2\right)-\frac{1}{2}\left(F_{\mu\rho}F_{\nu}^{\ \rho}-\frac{g_{\mu\nu}}{4}F_{\rho\sigma}F^{\rho\sigma}\right)\nonumber\\
&&+ \alpha \bigg[\phi H_{\mu\nu}  -2 R \left[(\nabla_{\mu} \phi)(\nabla_{\nu} \phi) + \nabla_{\nu} \nabla_{\mu} \phi \right] + 8 R_{({\mu}}^{\rho} \nabla_{{\nu})} \nabla_{\rho} \phi+ 8 R_{({\mu}}^{\rho} (\nabla_{{\nu})}\phi) (\nabla_{\rho} \phi)\nonumber\\
&& - 2 G_{\mu\nu} \left[(\nabla \phi)^2 +  2\square \phi \right] - 4 \left[ (\nabla_{\mu} \phi)(\nabla_{\nu} \phi) + \nabla_{\nu} \nabla_{\mu} \phi \right] \square \phi - \nonumber\\
&&\left[g_{\mu\nu}(\nabla \phi)^2 -4(\nabla_{\mu} \phi)(\nabla_{\nu} \phi) \right](\nabla \phi)^2
+ 8 (\nabla_{({\mu}} \phi) (\nabla_{{\nu})} \nabla_{\rho} \phi ) \nabla^{\rho} \phi\nonumber\\
&&- 4 g_{\mu\nu} R^{\rho\sigma} [\nabla_{\rho} \nabla_{\sigma} \phi  
+ (\nabla_{\rho} \phi)(\nabla_{\sigma} \phi) ] + 2 g_{\mu\nu} (\square \phi)^2- 2 g_{\mu\nu} (\nabla_{\rho} \nabla_{\sigma} \phi)(\nabla^{\rho} \nabla^{\sigma} \phi) \nonumber\\
&&- 4 g_{\mu\nu} (\nabla^{\rho} \phi ) (\nabla^{\sigma} \phi) (\nabla_{\rho} \nabla_{\sigma} \phi)+ 4 (\nabla_{\rho} \nabla_{\nu} \phi)(\nabla^{\rho} \nabla_{\mu} \phi)\nonumber\\
&&+ 4 R_{\mu\rho\nu\sigma} [(\nabla^{\rho} \phi)(\nabla^{\sigma} \phi)  + \nabla^{\sigma} \nabla^{\rho} \phi ]
+ 3 \lambda^2 e^{-4 \phi} g_{\mu\nu}\nonumber\\
&&- 2 \lambda e^{-2 \phi} \left( G_{\mu\nu} + 2 (\nabla_{\mu} \phi)(\nabla_{\nu} \phi) + 2 \nabla_{\nu} \nabla_{\mu} \phi - 2 g_{\mu\nu} \square \phi + g_{\mu\nu} (\nabla \phi)^2 \right)  \bigg]=0.
\end{eqnarray}
Combining the last two equations by $g^{\mu\nu}\mathcal{E}_{\mu\nu}+\alpha\mathcal{E}_{\phi}/2$, one obtains an equation independent of $\phi$:
\begin{equation}\label{eom}
g^{\mu\nu}\mathcal{E}_{\mu\nu}+\frac{\alpha}{2}\mathcal{E}_{\phi}=
4\Lambda-R-\frac{\alpha}2\mathcal{L}_{GB}+\frac 12\sum_{i=1}^{2}(\partial\varphi_i)^2=0.
\end{equation}
We assume that $\phi=\phi(r)$, and apply planar symmetric ansatz
\begin{equation}\label{metric}
ds_4^2=-e^{-2\chi(r)}f(r)dt^2+\frac 1{f(r)}dr^2+r^2(dx^2+dy^2).
\end{equation}
The scalar fields are linearly dependent on spatial coordinates as mentioned above, where
\begin{equation}
\varphi_i=\beta x_i.
\end{equation}
Now combining (\ref{eom}) together with (\ref{metric}), one obtains
\begin{eqnarray}\label{eom1}
&&-\frac{2 \alpha  \left(f'(r)^2+f(r) f''(r)\right)}{r^2}+\frac{4 f'(r)}{r}\nonumber\\
&&+f''(r)+\frac{2 f(r)}{r^2}+4 \Lambda +\frac{\beta ^2}{r^2}=0.
\end{eqnarray}

This equation is, however, not enough for us to find the explicit form of $f(r)$.  Substituting the metric (\ref{metric}) into (\ref{action}), and removing total derivative terms, one finds the effective Lagrangian to be \cite{2003}
\begin{eqnarray}\label{lag}
\mathcal{L}&=&\frac{e^{\chi(r)}}{6r^2}\left[-3 \left(4 r^3 f'(r)+h^2-q^2 e^{2 \chi (r)}+2 \beta ^2 r^2+4 \Lambda  r^4\right)\right.\nonumber\\
&&-4 r^2 f(r) \left(2 \alpha  f'(r) \phi '(r) \left(r^2 \phi '(r)^2-3 r \phi '(r)+3\right)+3\right)\nonumber\\
&&+4 \alpha  r^2 f(r)^2 \phi '(r) (4 \chi '(r) \left(r^2 \phi '(r)^2-3 r \phi '(r)+3\right)\nonumber\\
&&\left.\left.+\phi '(r) \left(3 r^2 \phi '(r)^2-8 r \phi '(r)+6\right)\right)\right],
\end{eqnarray}
where flat ``internal'' space is considered, which means $\lambda =0$ and the theory is invariant under a constant shift of $\phi$.\\
Taking variation of (\ref{lag}) with respect to $f(r)$, and making $\chi(r)$ to be zero, one finds
\begin{equation}
4 \alpha  f(r) \left(r \phi '(r)-1\right)^2 \left(\phi '(r)^2+\phi ''(r)\right)=0,
\end{equation}
which implies
\begin{equation}\label{phi}
\phi '(r)= \frac{1}{r},
\end{equation}
and inserting (\ref{phi}) into $\delta \chi$ equation yields
\begin{eqnarray}\label{eom2}
&&4 r \left(r^2-2 \alpha  f(r)\right) f'(r)+4 f(r) \left(\alpha  f(r)+r^2\right)\nonumber\\
&&+h^2+q^2+2 \beta ^2 r^2+4 \Lambda  r^4=0.
\end{eqnarray}
Let $r_H$ to be the black brane horizon, i.e. $f(r_H)=0$. One is now able to give the exact solution from (\ref{eom1}) and (\ref{eom2}):
\begin{eqnarray}
f(r)&=&\frac{r^2}{2\alpha }\left(1-\left(1+\frac{4 \alpha  \Lambda }{3}-\frac{4 \alpha  \Lambda  r_H^3}{3 r^3}-\frac{\alpha  h^2}{r^4}+\frac{\alpha  h^2}{r^3 r_H}\right.\right.\nonumber\\
&&\left.\left.-\frac{\alpha  q^2}{r^4}+\frac{\alpha  q^2}{r^3 r_H}+\frac{2 \alpha  \beta ^2}{r^2}-\frac{2 \alpha  \beta ^2 r_H}{r^3}\right)^{\frac 12}\right).
\end{eqnarray}

\subsection{Planar black brane in AdS space}
Now rescale the parametres, with re-definition as follows:
\begin{eqnarray}
\alpha_{GB}&=&-\frac{\Lambda}{3}\alpha,\\
\hat{\beta}^2&=&-\frac 3{2\Lambda}\frac {\beta^2}{r_H^2},\\
Q^2&=&-\frac 3{4\Lambda}\frac{q^2}{r_H^4},\\
H^2&=&-\frac 3{4\Lambda}\frac{h^2}{r_H^4}.
\end{eqnarray}
The line element for planar black brane in AdS space can be written as
\begin{equation}\label{mef}
ds^2=-F(r)N^2dt^2+F(r)^{-1}dr^2+\frac{r^2}{l^2}(dx^2+dy^2),
\end{equation}
with $l$ the AdS radius, and
\begin{eqnarray}
F(r)&=&\frac{r^2}{2\alpha_{GB}l^2}\left(1-\left(1-4\alpha_{GB}\left(1-\frac{r_H^3}{r^3}(1+H^2+Q^2)\right.\right.\right.\nonumber\\
&&\left.\left.\left.+\frac{r_H^4}{r^4}(H^2+Q^2)+\hat{\beta}^2(\frac{r_H^3}{r^3}-\frac{r_H^2}{r^2})\right)\right)^{\frac 12}\right),
\end{eqnarray}
where we have used the fact that when $D=4$,
\begin{equation}
\Lambda=-\frac{(D-1)(D-2)}{2l^2}=-\frac{3}{l^2}.
\end{equation}
To find the value of constant $N^2$, one needs to note that the geometry would reduce to the flat Minkowski metric conformally at the boundary. With $r \to \infty$, one has
\begin{equation}
F(r)N^2\to \frac{r^2}{l^2}.
\end{equation}
As a result,
\begin{equation}
N^2=\frac 12(1+\sqrt{1-4\alpha_{GB}}).
\end{equation}
This solution is a dyonic black hole in $4D$ EMGB gravity with linear axions. Charged black holes in four dimensions have been studied for long time, and many interesting properties of them in transport and thermodynamic behaviour have been found \cite{charge1,charge2,charge3}. This is the first time for a dyonic black hole coupled with Gauss-Bonnet terms thanks to the ``Kaluza-Klein-like'' method.

Since Gauss-Bonnet term only contains curvature terms, the black hole thermodynamics is the same as that of Schwarzchild-AdS. This property is preserved by our reduction frame.
The Hawking temperature at the event horizon reads
\begin{eqnarray}\label{temp}
T&=&\frac{1}{2\pi\sqrt{g_{rr}}}\frac{d \sqrt{g_{tt}}}{d r}|_{r=r_H}\nonumber\\
&=&\frac{N}{4\pi}F'(r_H)
=\frac{Nr_H}{4\pi l^2}(3-\hat{\beta}^2-H^2-Q^2).
\end{eqnarray}
The black brane approaches extremal when $T \to 0$, that is,
\begin{equation}
\hat{\beta}^2+H^2+Q^2=3.
\end{equation}
The entropy density of the horizon is given by \cite{v2,v3}
\begin{equation}
s=\frac{r_H^2}{4G_4l^2}.
\end{equation}
At $T=0$, one has $\hat{\beta}^2=3-H^2-Q^2$. We can expand $F(r)$ around $r\sim r_H$ to investigate the geometry near horizon. One obtains
\begin{equation}
F(r)\simeq\frac{3+H^2+Q^2}{l^2}(r-r_H)^2+\mathcal{O}\left((r-r_H)^3\right).
\end{equation}
By setting 
\begin{equation}
r-r_H=\frac{l^2}{3+H^2+Q^2}\frac{1}{\bar{r}},
\end{equation}
one finds that the near-horizon metric at zero temperature reads
\begin{equation}
ds^2=\frac{\mathfrak{L}^2}{\bar{r}^2}(-dt^2+d\bar{r}^2)+\frac{r_H}{l^2}(dx^2+dy^2),
\end{equation}
where $\mathfrak{L}$ is the curvature radius of $AdS_2$ that takes the form
\begin{equation}
\mathfrak{L}=\sqrt{\frac{l^2}{3+H^2+Q^2}}.
\end{equation}
This implies that the extremal black brane geometry is equivalent to $AdS_2\times\mathbb{R}^2$ topologically.

Having derived the new black hole solution, we will continue to explore more on its  properties. The shear viscosity to entropy density ratio is going to be investigated in the next section. One has three parametres in total: the electric charge $Q$, the magnetic charge $H$ and the scalar field constant $\hat{\beta}$, making the calculation we will perform too complicated to be done. Nevertheless, we do not compute the electric conductivity in this paper. What makes the black hole solution in this paper really new is the coupling of scalar fields and Gauss-Bonnet terms. To study the new system step by step, it will be a good choice to first only consider the contributions from scalar fields. Moreover, it has been proved that the bound can be violated when the black hole is charged, and the charge does not influence the causal structure \cite{4d1,v3}. For these reasons, we can remove the charges and calculate the ratio, after which the behaviour with non-vanishing charges can be analysed qualitatively without significant error.

\section{Shear viscosity for neutral black holes}
\subsection{Weaker horizon formula}
From now on, only black holes without electric or magnetic charge are considered, i.e. $H=Q=0$. Before the shear viscosity is investigated, let us change the coordinates where $u=r_H/r$, and
\begin{equation}
f(u)=\frac{1}{2\alpha_{GB}}\left(1-\sqrt{1-4\alpha_{GB}\left(1-u^2+\hat{\beta}^2(u^2-u^3)\right)}\right).
\end{equation}

The tensor type perturbation is considered where
\begin{equation}\label{pert}
(\delta g)^x_y=h(t,u)=h(u)e^{-i\omega t},
\end{equation}
and
\begin{eqnarray}\label{metricpert}
ds^2&=&\frac{r_H^2}{u^2 l^2}\left(-f(u)N^2dt^2+d\vec{x}^2+2h(t,u)dxdy\right)+\frac {l^2}{u^2f(u)}du^2.
\end{eqnarray}
Usually Kubo formula gives the shear viscosity in general cases:
\begin{equation}
\eta=\lim\limits_{\omega\to 0}\frac{1}{2\omega}\int dtd\mathbf{x}e^{i\omega t}\langle[T_{xy}(t,\mathbf{x}),T_{xy}(t,\mathbf{0})]\rangle,
\end{equation}
where $T_{xy}$ is the momentum-stress tensor. If translation invariance is not broken by any sources, then the momentum $T_{ty}$ is conserved, whose current corresponds to $T_{xy}$. The KSS bound is obtained in momentum-conserving situation. It is weakened and becomes (\ref{gbbound}) when higher derivative terms are non-vanishing.

Now with scalar fields, the momentum is no longer conserved, and the translation invariance is broken. As a consequence, the shear viscosity $\eta$ does not have a hydrodynamic interpretation any more, but $\eta/s$ is still closely associated with the entropy production. In this situation, one cannot simply apply Kubo formula to compute shear viscosity $\eta$, and it will not work if one tries to find it only via horizon data directly. Instead, a ``weaker horizon formula'' has been suggested to study shear viscosity to entropy density ratio $\eta/s$ in such cases \cite{weak}.  While the perturbation is massive, one has \cite{weak}
\begin{equation}\label{weaker}
\frac{\eta}{s}=\frac{1}{4\pi}h(1)^2.
\end{equation}
Substituting (\ref{metricpert}) into (\ref{eom}) and taking $\omega\to0$, one obtains
\begin{eqnarray}\label{eompert}
&&-4 h(u) [f(u) \left(h'(u) (\alpha_{GB}  u^2 f''(u)-6 \alpha_{GB}  u f'(u)-3)+u h''(u) (\alpha_{GB}  u f'(u)+1)\right)\nonumber\\
&&+u f'(u) h'(u) \left(\alpha_{GB}  u f'(u)+1\right)+f(u)^2 (6 \alpha_{GB}  h'(u)-2 \alpha_{GB}  u h''(u))]\nonumber\\
&&+u h'(u) (f(u) \left(2 \alpha_{GB}  u^2 f'(u) h''(u)+h'(u) (\alpha_{GB}  u^2 f''(u)-8 \alpha_{GB}  u f'(u)-3)\right)\nonumber\\
&&+\alpha_{GB}  u^2 f'(u)^2 h'(u)+2 \alpha_{GB}  f(u)^2 \left(5 h'(u)-2 u h''(u)\right))+4 \hat{\beta}^2  u h(u)^2=0,
\end{eqnarray}
with boundary conditions
\begin{eqnarray}
h(0)&=&\sqrt{1-6\alpha_{GB}},\label{bc}\\
h(1)&=&{regular}.
\end{eqnarray}

We rewrite the Hawking temperature (\ref{temp}) as
\begin{eqnarray}\label{temp2}
T&=&\frac{N r_H}{4\pi l^2}3\hat{\beta}(\frac{1}{\hat{\beta}}-\frac{\hat{\beta}}{3})=\frac{3}{4\sqrt{2}}\frac{N\beta}{ l}\frac{1}{\pi}(\frac{1}{\hat{\beta}}-\frac{\hat{\beta}}{3})=\xi\frac{1}{\pi}(\frac{1}{\hat{\beta}}-\frac{\hat{\beta}}{3}),
\end{eqnarray}
where
\begin{equation}
\xi=\frac{3}{4\sqrt{2}}\frac{N\beta}{ l}.
\end{equation}

As (\ref{eompert}) is still too complicated to be solved neither analytically nor numerically, the solution will be found at high and low temperatures separately.

\subsection{High temperature expansion}
At high temperature, $\hat{\beta}^2\to0$. Therefore, (\ref{eompert}) could be perturbatively solved around $\hat{\beta}^2\sim 0$. Along this line, $h(u)$ will be expanded as
\begin{equation}
h(u)=\sum_{i=0}^{n}\hat{\beta}^{2i}h_{2i}(u).
\end{equation}

At the $0$th order, $h_0(u)$ turns out to be a constant function. With the boundary condition (\ref{bc}), one has
\begin{equation}
h_0(u)=\sqrt{1-6\alpha_{GB}}.
\end{equation}

At the second order, one finds
\begin{eqnarray}\label{eomh}
&&\frac{1}{\alpha_{GB}  \left(4 \alpha_{GB}  \left(u^3-1\right)+1\right)^{3/2}}[2 \sqrt{1-4 \alpha_{GB} } [3 (2 \alpha_{GB} ^2 \left(4 u^3 \left(2 \sqrt{4 \alpha_{GB}  \left(u^3-1\right)+1}-3\right)\right.\nonumber\\
&&\left.-8 \sqrt{4 \alpha_{GB}  \left(u^3-1\right)+1}+u^6+8\right)+\alpha_{GB}  \left(u^3 \left(6-4 \sqrt{4 \alpha_{GB}  \left(u^3-1\right)+1}\right)\right.\nonumber\\
&&\left.+8 \left(\sqrt{4 \alpha  \left(u^3-1\right)+1}-1\right)\right)-\sqrt{4 \alpha_{GB}  \left(u^3-1\right)+1}+1) {h_2}'(u)+u (4 \alpha_{GB} ^2 \left(u^6-5 u^3+4\right)\nonumber\\
&&+\alpha_{GB}  \left(5 u^3-8\right)+1) \left(\sqrt{4 \alpha_{GB}  \left(u^3-1\right)+1}-1\right) {h_2}''(u)]]+4 (1-4 \alpha_{GB} ) u=0.
\end{eqnarray}

At this order, one has
\begin{equation}
\frac{\eta}{s}=h_0(1)^2+2h_0(1)h_2(1)\hat{\beta}^2+\mathcal{O}(\hat{\beta}^3).
\end{equation}

Also,
\begin{equation}
\frac{T}{\xi}\simeq\frac{1}{\pi}\frac{1}{\hat{\beta}},
\end{equation}
so
\begin{equation}
\hat{\beta}\simeq\pi\frac{\xi}{T}.
\end{equation}

\begin{figure}[tbp]
	\centering 
	\includegraphics[width=0.7\textwidth]{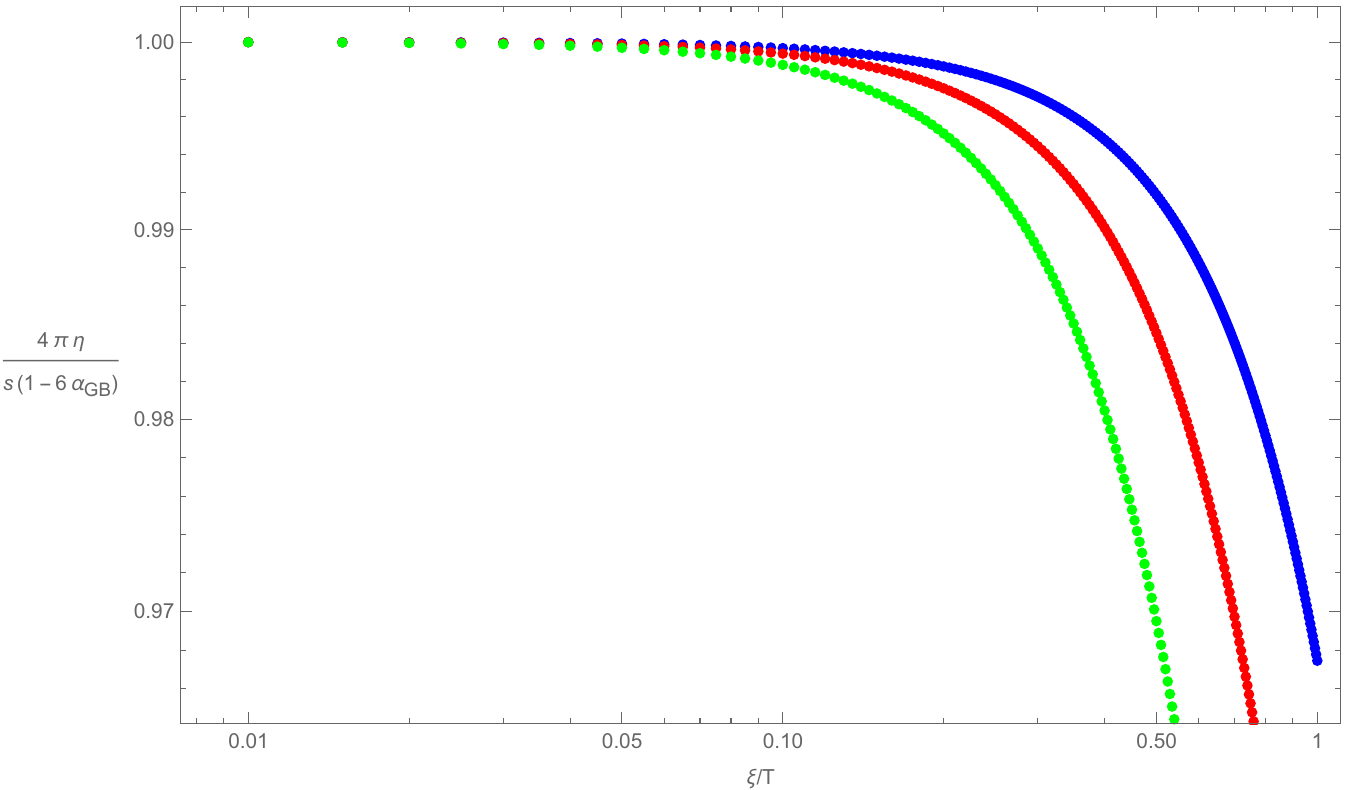}
	\hfill
	\caption{\label{hightemp}Log-log plot of numerical solution for $\eta/s$ at high temperature as a function of $\xi/T$. From bottom to top, the green, red and blue lines represents solutions for $\alpha_{GB}=0.15, 0.1, -0.1$ respectively..}
\end{figure}

The numerical solution for (\ref{eomh}) is found and the ratio $\eta/s$ as a function of $\xi/T$ is illustrated, which is shown in Fig.(\ref{hightemp}) as a log-log plot. One finds that at high temperature, the bound (\ref{gbbound}) is violated when the temperature is getting lower.

\subsection{Low temperature expansion}

With (\ref{temp2}), one has $\hat{\beta}^2\sim3$ at low temperature. Similarly,
\begin{equation}
h(u)=\sum_{i=0}^{\infty}(\hat{\beta}^2-3)^ih_i(u).
\end{equation}

As the equations of motion are rather complicated, and what is needed is only the value of $h(1)$, we are not going to give the explicit form of the equations in this paper. However, at the $0$th order, one finds that
\begin{eqnarray}
h_0'(0)&=&0,\\
h_0(1)&=&0,
\end{eqnarray}
and at the first order, one obtains
\begin{eqnarray}
h_1'(0)&=&0,\\
h_1(1)&=&\frac 16 h_0'(1).
\end{eqnarray}

According to (\ref{weaker}), to the first order, the shear viscosity to entropy density ratio at low temperatures reads
\begin{eqnarray}
4\pi\frac{\eta}{s}&=&h_0(1)^2+2 (\hat{\beta}^2 -3) h_0(1) h_1(1)+(\hat{\beta}^2 -3)^2 h_1(1)^2+\mathcal{O}((\hat{\beta}^2 -3)^3)\nonumber\\
&\simeq&(\hat{\beta}^2 -3)^2 h_1(1)^2=\frac{1}{36}(\hat{\beta}^2 -3)^2h_0'(1)^2.
\end{eqnarray}

Therefore, the problem reduces to finding out the value of $h_0'(1)$. Though the equation is too cumbersome to be solved directly, one could follow the similar method performed in the previous work, solving $h_0(u)$ at $u=0$ and $u=1$ respectively, and matching the solutions to obtain $h_0'(1)$ \cite{v5}. Here is the strategy:
\begin{itemize}
	\item [1)]
	Solve $h_0(u)$ at $u=0$ and $u=1$, and these two solutions are labeled as $h_{00}(u)$ and $h_{01}(u)$ respectively. At this stage, the integral constant in $h_{01}(u)$ is still not fixed.
	\item[2)]
	Assume $h_{00}(u)$ and $h_{01}$ match at $u=u_m$, where
	\begin{equation}\label{condition1}
	h_{00}(u_m)=h_{01}(u_m),
	\end{equation}
	and
	\begin{equation}\label{condition2}
	h_{00}'(u_m)=h_{01}'(u_m).
	\end{equation}
	\item[3)]
	Using the conditions above, one could fix $h_{01}(u)$, thus find $h_{01}'(1)$.
\end{itemize}

Following the steps above, one finds that at $u=0$,
\begin{equation}
h_{00}(u)=\sqrt{1-6\alpha_{GB}}.
\end{equation}

Similarly, at the horizon where $u=1$,
\begin{equation}
h_{01}''(1)= \frac{(26 \alpha +5) h_{01}'(1)}{6 (8 \alpha -3)},
\end{equation}
so
\begin{eqnarray}
h_{01}(u)&=&h_{01}(1)+(u-1) h_{01}'(1)+\frac{1}{2} (u-1)^2 h_{01}''(1)+\mathcal{O}\left((u-1)^3\right)\nonumber\\
&\simeq&\frac{1}{96 \alpha_{GB} -36}[(u-1) (70 \alpha_{GB} +(26 \alpha_{GB} +5) u-41)] h_{01}'(1).
\end{eqnarray}
With (\ref{condition1}) and (\ref{condition2}), one obtains
\begin{equation}
h_{01}'(1)=\frac{\sqrt{1-6 \alpha_{GB} } (26 \alpha_{GB} +5)}{9-24 \alpha_{GB} }.
\end{equation}

Next thing to do is exactly the same as that in high-temperature case: write $\eta/s$ as a function of $\xi/T$ according to (\ref{temp2}). As $\hat{\beta}^2\to3$,
\begin{equation}
\frac{T}{\xi}=-\frac{1}{3\pi\hat{\beta}}(\hat{\beta}^2-3)\simeq-\frac{1}{3\sqrt{3}\pi}(\hat{\beta}^2-3).
\end{equation}

Therefore, at low temperature,
\begin{equation}\label{lowsol}
4\pi\frac{\eta}{s}=\frac{1}{36}\left(3\sqrt{3}\pi\frac{T}{\xi}\right)^2\left(\frac{\sqrt{1-6 \alpha_{GB} } (26 \alpha_{GB} +5)}{9-24 \alpha_{GB} }\right)^2.
\end{equation}

\begin{figure}[tbp]
	\centering 
	\includegraphics[width=0.7\textwidth]{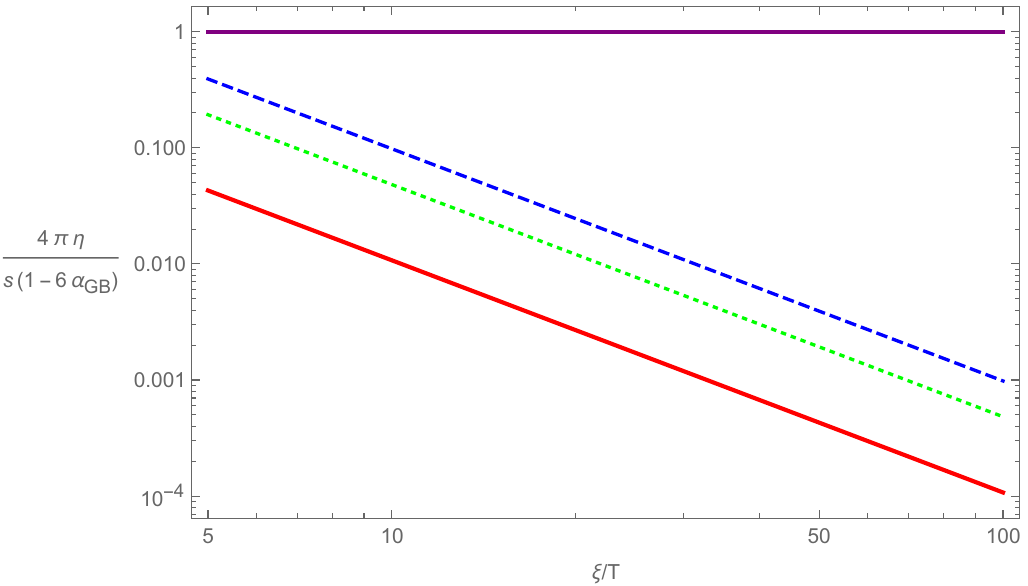}
	\hfill
	\caption{\label{lowtemp}Log-log plot for $\eta/s$ at low temperature as a function of $\xi/T$. From top to bottom, the blue dashed line stands for $\alpha_{GB}=0.1$, and the green dotted line represents $\alpha_{GB}=0.05$, while the red thick line for $\alpha_{GB}=-0.5$.}
\end{figure}

One finds from Fig.(\ref{lowtemp}) that the KSS bound is violated as well.

In spite of the fact that ``Kaluza-Klein-like'' process gives us really different equations of motion from those in $5D$ spacetime, the behaviour $\eta/s$ is quite similar with what has been found in five dimensions \cite{v4}. In our case, $\eta/s\sim(T/\xi)^2$ when $T/\xi \to 0$, which satisfies the conjecture \cite{v1} that $\eta/s\sim(T/\Delta)^2$ as $T/\Delta \to 0$ with $\Delta$ being some scale to be chosen. Here we take $\Delta$ to be $\xi$.

\subsection{Discussion}
It is obvious that at high temperature, the KSS bound in higher derivative gravity $\eta/s\geq (1-6\alpha_{GB})/4\pi$ is hardly violated. This is because the high temperatures correspond to very small $\hat{\beta}^2$, where the contribution from axions is almost neglectable. While the low-temperature behaviour of the ratio violates the bound dramatically, since in this situation, $\hat{\beta}^2$ is a large number comparing with that in high-temperature cases. Thus, for fixed Gauss-Bonnet constant, the larger the mass of the graviton, the bolder the violation would be, which evinces one's intuitive postulation. 

One may find that (\ref{lowsol}) looks rather different with that in $5$ dimensions \cite{v5}, who contains the confluent hypergeometric limit function related with Bessel functions. This results from the fact that our theory in four-dimensional spacetime is different from those in higher dimensions, since it also includes the ``breathing scalar'' $\phi$ that yields a completely different Einstein equations. All calculation is based on these equations of motion, so it is natural that in this paper one would get a quite different form of the ratio. 

Nevertheless, the critical characteristics of $\eta/s$ is almost the same in $4-$ and $5-$ dimensional spacetime. They both violates KSS bound markedly. Furthermore, in both cases, $\eta/s\sim(T/\Delta)^2$, and $\Delta\sim\beta$, only differing by a coefficient that is determined by the dimensionality. Although governed by different actions and therefore different equations, the main features characterising the shear viscosity to entropy density ratios in four and higher dimensions are actually very similar to each other.

In the situation where the charges $Q$ and $H$ are non-zero, one has
\begin{eqnarray}
f(u)&=&\frac{1}{2\alpha_{GB}}\left(1-\left(1-4\alpha_{GB}\left(1-u^2(1+H^2+Q^2)\right.\right.\right.\nonumber\\
&&\left.\left.\left.+u^4(H^2+Q^2)+\hat{\beta}^2(u^2-u^3)\right)\right)^{\frac{1}{2}}\right).
\end{eqnarray}

The content above has shown the complicatedness of obtaining the ratio $\eta/s$ even one only has dropped the electric and magnetic charges away. The equation involving $H$ and $Q$ is beyond the scope of this paper. However, one may analyse the possible outcome for the full solution from the investigation on neutral black holes with axions and the charged ones qualitatively.\\
For charged black holes without axions in $4D$ EGB gravity, one has \cite{4d1},
\begin{eqnarray}
\frac{\eta}{s}&=&\frac{1}{4\pi}\left(1-2\alpha_{GB}\frac{(D-1)-(D-3)(H^2+Q^2)}{D-3}\right)\nonumber\\
&=&\frac{1}{4\pi}\left[1-6\alpha_{GB}\left(1-\frac{1}{3}(H^2+Q^2)\right)\right],
\end{eqnarray} 
which certainly breaks the KSS bound for negative $\alpha_{GB}$. It is worth noting that there exists a constraint on $\alpha_{GB}$ that one should choose $\alpha_{GB}<0$ when the black hole is charged, because of the causality \cite{4d1} and the completeness of the spacetime \cite{miao}. Hence, the ratio $\eta/s$ of charged black holes will be definitely violated.\\
Considering that the existence of the scalar fields break the KSS bound as well, if we add the charges back, it is reasonable to expect that the KSS bound should still be violated.

\section{Causality}
\subsection{Inability of ``Kaluza-Klein-like'' process in causality analysis}
When introducing Gauss-Bonnet terms, one would find that the causality will be violated, and charge as well as scalar fields have effects on the violation \cite{v1,v2,v3,v4,v5}. Through analysis of causal structure, one is capable of finding restrictions on $\alpha_{GB}$. For example, in $5$ dimensions, causality will be violated if $\alpha_{GB}>0.09$ \cite{v1,v2} with non-vanishing charges. 

We would like to study the causal structure of the bulk. According to the AdS/CFT correspondence, $4D$ AdS gravity is dual to a $3D$ quantum field theory on the boundary.

Usually, the procedure to study the causality in dimension $D$ is: 

\begin{itemize}
	\item [1)] 
	Start with a $D$-dimensional metric
	\begin{eqnarray}\label{metric2}
	ds^2&=&-N^2f(r)dt^2+\frac{1}{f(r)}dr^2+\frac{r^2}{l^2}\left(\sum_{a=1}^{D-2}dx_a^2+2h(x_i,r)dx_mdx_n\right).
	\end{eqnarray}
	Write the perturbation (which is the wave function of the transverse graviton) as
	\begin{equation}\label{wf}
	h(x_i,r)=e^{-i\omega t+ik_r r+ikx_i}.
	\end{equation}       
	\item [2)]
	Then take large momentum limit, where $k^{\mu}\to\infty$. The $x_m x_n$-component of equation of motion will reduce to 
	\begin{equation}\label{kt}
	k^{\mu}k^{\nu}g_{\mu\nu}^{{eff}}\simeq 0,
	\end{equation}
	where
	\begin{eqnarray}
	ds^2_{{eff}}&=&g_{\mu\nu}dx^{\mu}dx^{\nu}\nonumber\\
	&=&N^2f(r)\left(-dt^2+\frac{1}{c_g^2}dx_i^2\right)+\frac{1}{f(r)}dr^2
	\end{eqnarray}
	is the effective metric.
	\item [3)]
	Find $c_g$ and the constraint by requiring $c_g^2-1\le 0$.
\end{itemize}
It is important to mention that $c_g$ can be interpreted as the local speed of graviton on a constant $r$-hypersurface. Its dependence on dimensionality is given by \cite{v4}
\begin{eqnarray}\label{cg}
c_g^2(\mathfrak{r})&=&\frac{N^2 f}{\mathfrak{r}^2}\frac{1}
{1-\frac{2\mathfrak{a}}{D-3}\left(\mathfrak{r}^{-1}f'+\mathfrak{r}^{-2}(D-5)f\right)}\nonumber\\
&&[1-\frac{2\mathfrak{a}}{(D-3)(D-4)}\left(f''+(D-5)(D-6)\mathfrak{r}^{-2}f+2(D-5)\mathfrak{r}^{-1}f'\right)],
\end{eqnarray}
where $\mathfrak{r}=r/r_H$, and $\mathfrak{a}=(D-3)(D-4)\tilde{\alpha}/l^2$. The local speed of light is defined as $c_b=N^2f(\mathfrak{r})/\mathfrak{r}^2$. At the boundary, where $\mathfrak{r}\to \infty$, $c_b$ is $1$. It seems that if we simply rescale the Gauss-Bonnet constant $\tilde{\alpha}$ into $\alpha/(D-4)$, then we can get rid of $(D-4)$ term, and find the constraint on $\mathfrak{a}$ by taking $\mathfrak{r}\to \infty$ and $D\to 4$.

Before doing it, let us come back to the ``Kaluza-Klein-like'' procedure \cite{2003,2004} applied in this paper. It is expected that we could also make a similar perturbation like (\ref{wf}), and directly find the momentum term from (\ref{eom}), or just from $\mathcal{E}_{\mu\nu}$.

Since we only have $4$ dimensions, one should either take $x_i$ in (\ref{wf}) to be $x$ or $y$. As a consequence, the momentum $k$ will be in $x$ (or $y$) direction. More precisely, for instance, one has 
\begin{equation}\label{px}
h(x,u)=e^{-i\omega t+ik_u u+ikx}.
\end{equation}
However, if one tries to substitute (\ref{px}) into (\ref{eom}), no momentum term could be found, neither in  (\ref{eom}) nor in $\mathcal{E}_{\mu\nu}$. This implies that it is impossible to analyse causal structure through ``Kaluza-Klein-like'' procedure as far as we are concerned currently. 

\subsection{Vanishing momentum terms in four dimensions}
Turning back to (\ref{cg}), on the other hand, one is able to study the causality with this formula at the first sight. But one has to be careful when performing such a limitation.

There is a rather simple way to express (\ref{pert}) more generally, which also tells why the situation becomes more complicated when we deal with $4$ dimensions.

It will be useful to rewrite the metric (\ref{metric2}) as
\begin{eqnarray}\label{metric3}
ds^2&=&-N^2f(r)dt^2+\frac{1}{f(r)}dr^2+\frac{r^2}{l^2}(dx^2+dy^2+2h(z_a,r)dxdy)+\mathfrak{z}\frac{r^2}{l^2}
\sum_{i=1}^{D-4}dz_i^2,
\end{eqnarray}
where $\mathfrak{z}=1$ when $D\geq 5$, and $\mathfrak{z}=0$ for $D\leq 4$. With (\ref{metric3}), one obtains 
\begin{eqnarray}\label{cg2}
c_{g,\mathfrak{z}}^2(\mathfrak{r})&=&\frac{1}{\mathfrak{z}}\frac{N^2 f}{\mathfrak{r}^2}\frac{1}
{1-\frac{2\mathfrak{a}}{D-3}\left(\mathfrak{r}^{-1}f'+\mathfrak{r}^{-2}(D-5)f\right)}\nonumber\\
&&[1-\frac{2\mathfrak{a}}{(D-3)(D-4)}\left(f''+(D-5)(D-6)\mathfrak{r}^{-2}f+2(D-5)\mathfrak{r}^{-1}f'\right)].
\end{eqnarray}
From (\ref{cg2}), one finds that if one works with dimensions higher than $4$, then $\mathfrak{z}=1$, and she or he will definitely recover (\ref{cg}). But if we take $D=4$, we should also make $\mathfrak{z}=0$ at the same time, where (\ref{cg}) works no more. What will be found is an infinite graviton velocity. In short,
\begin{equation}
\left.c_g^2\right|_{D=4}\neq\lim\limits_{D\to 4}c_g^2.
\end{equation}

Furthermore, the momentum term in $D$-dimensional equation of motion reads $c_{g,\mathfrak{z}}^2k^2/(N^2f(\mathfrak{r}))$. Obviously, $\mathfrak{z}$ appears in denominator. When $D$ is $4$, $k$ has to vanish in order that the momentum term would not diverge.

One could turn back to the very beginning to see why $4D$ cases are so special. The answer is quite simple: the momentum term containing $k$ vanishes if $x_i=x_m$ or $x_i=x_n$, in spite of how many dimensions one has.

Consequently, at least three spatial coordinates, i.e. five dimensions in total are required to construct the perturbation in (\ref{metric2}) that leads to non-zero momentum.  For example, in $5$-dimensional spacetime, we often choose $x_mx_n$ to be $xy$, and thus $x_i$ is $z$.

However, the statement above is only for the cases of the black holes with axions obtained by ``Kaluza-Klein-like'' procedure. One may find a different result with other choices. For charged black holes in $4D$ EGB gravity with various methods, there in a constraint such that $\alpha_{GB}<0$ \cite{4d1,miao}. 

Causal analysis is not the only way to get this result \cite{4d1}, and investigation into the completeness of the spacetime also yields a similar upshot \cite{miao}. Hence, it is reasonable to expect that based on ``Kaluza-Klein-like'' method, one could get a similar constraint on $\alpha_{GB}$ too.

\section{Conclusion}
In this paper, we obtained the dyonic black hole solution with linear axions in $4$-dimensional higher derivative gravity through ``Kaluza-Klein-like'' process. The shear viscosity to entropy density ratio $\eta/s$ is investigated after electric and magnetic charges are removed. It turns out that violation still happens when $D=4$. 

The behaviour of $\eta/s$ is rather similar with that in $5$ dimensions, such that $\eta/s\sim(T/\xi)^2$ when $T/\xi$ is very small. One important outcome is that the main feature of the ratio is almost the same with what has been found in $5$ dimensions \cite{v5}. The only difference comes from the different equations of motions brought by the ``breathing scalar'', which is inevitable if one applies ``Kaluza-Klein-like'' process to get the four-dimensional theories. With this result, we suppose that the KSS should also be violated if the charges are non-zero. 

While the bulk causal structure of the new black holes is studied, it is found that the momentum term vanishes in equations of motion, while the velocity formula in $D$ dimensions is only valid when $D>4$. Therefore, ``Kaluza-Klein-like'' process cannot be applied on causality investigation, since the construction of causality analysis is only meaningful in dimensions no lower than $5$. 

As is mentioned, it has been shown that charged black holes in $4D$ EGB gravity from different processes have shown a constraint on the Gauss-Bonnet constant. Since we only consider neutral black hole with tensor type perturbations in this paper, our next task may focus on the full solution with charges through a new method instead of causal analysis. Different types of perturbations are of our future research interest as well. The research work done previously implies a possibility that the bound on Gauss-Bonnet constant could be found. We expect a constraint similar with $\alpha_{GB}<0$ for charged black holes obtained from ``Kaluza-Klein-like'' procedure. 

The dyonic black hole solution derived may be used as a tool to study the transport properties of the normal state of high-temperature superconductors. It is of our further interest as well to go further and explore more of its properties, such as transport behaviour like electric and thermal conductivity. Much interesting upshot is expected to appear, as this is the first time for a $4$d dyonic black hole to contain contributions from higher-derivative terms, which may bring an insight into the study on high-temperature superconductivity.

\appendix

\acknowledgments
We would like to thank Chun-Shan Lin and Yan-Gang Miao for helpful discussions at the early stage of this work. This work is partly supported by NSFC, China (No.11875184).

\end{document}